\documentclass[11pt]{article}
\usepackage{latexsym}
\usepackage{amsmath,amsfonts,amssymb,enumerate,multicol,tikz}
\usepackage{epsfig}
\usepackage{pstricks}
\usepackage{hyperref}

\usepackage{mathrsfs}

\begin{document}
\newtheorem{definition}{Definition}[section]
\newtheorem{theorem}[definition]{Theorem}
\newtheorem{lemma}[definition]{Lemma}
\newtheorem{proposition}[definition]{Proposition}
\newtheorem{examples}[definition]{Examples}
\newtheorem{corollary}[definition]{Corollary}
\def\square{\Box}
\newtheorem{remark}[definition]{Remark}
\newtheorem{remarks}[definition]{Remarks}
\newtheorem{exercise}[definition]{Exercise}
\newtheorem{example}[definition]{Example}
\newtheorem{observation}[definition]{Observation}
\newtheorem{observations}[definition]{Observations}
\newtheorem{algorithm}[definition]{Algorithm}
\newtheorem{criterion}[definition]{Criterion}
\newtheorem{algcrit}[definition]{Algorithm and criterion}

\pagestyle{plain}

\newenvironment{prf}[1]{\trivlist
\item[\hskip \labelsep{\it
#1.\hspace*{.3em}}]}{~\hspace{\fill}~$\square$\endtrivlist}

\title{The geometric semantics of algebraic quantum mechanics} 
\author{John Alexander Cruz Morales\footnote{e-mail: alekosandro@gmail.com, jacruzm@impa.br}\\
Instituto Nacional de Matem\'{a}tica Pura e Aplicada, IMPA. \\
Estrada Dona Castorina 110. Rio de Janeiro, 22460-320, Brasil.\\
Boris Zilber\footnote{e-mail: zilber@maths.ox.ac.uk}\\
Mathematical Institute.\\ 
University of Oxford.\\
24-29 St Giles. Oxford, OX1 3LB, United Kingdom.}

\date{August 24 2014}

\maketitle

\begin{abstract} In this paper we will present an ongoing project which aims to use model theory as a suitable mathematical setting for studying the formalism of quantum mechanics. We will argue that this approach provides a geometric semantics for such formalism by means of establishing a (non-commutative) duality between certain algebraic and geometric objects. 
\end{abstract}

\section{Introduction}

The idea of using model theory as a mathematical framework for providing rigorous and solid foundations for (quantum) physics comes from the paper \cite{bo4} and later  preprints  of the second author. During these years some of the ideas in that work have maturated in the yet unpublished paper \cite{bo3}.
We give here an account of the current status of this work and trace some ideas for future developments. In particular, in this paper, we will focus on the formalism of quantum mechanics and will describe how to construct a geometric semantics for it by means of using model-theoretic tools. In fact, this project is a part of a much broader one which aims to find in a systematic way a geometric counterpart $\mathbf{V}_\mathcal{A}$ for a given non-commutative algebra $\mathcal{A}$ (which should be understood as a `coordinate algebra' in the sense of non-commutative geometry). Here ``geometric'' should also be understood in some broad but well-defined sense. \\

We can think of model theory as the mainstream mathematics which pays special attention to the language used and definability issues so, very roughly, what we want to achieve is to do physics paying special attention to the language and definability issues. This way of thinking might be helpful for addressing questions like:  \\
1. What structures do physicists use? \\
2. What kind of number system do physicists use? \\
3. What kind of limit procedures, approximations do physicists use? \\
In this text we are going to sketch a first approach for tackling these questions. \\

Model theory has worked out a very efficient hierarchy of type of structures (stability theory) and, in particular, an important class of structures was introduced by Hrushovski and the second author \cite{hrbo} in order to identify and characterise logically perfect structures, i.e., the top level of the stability hierarchy. These structures are called Zariski geometries and their links with non-commutative geometry and quantum physics is an essential part of the present paper via a (non-commutative) extension of the classical duality between algebraic and geometric objects. This duality may be reinterpreted in terms of the duality syntax/semantics and this is an idea we will have in mind in this paper.\\

From a mathematical point of view the main idea of our project consists in establishing a rigorous duality  

\begin{equation*}
\mathcal{A}_{\bf{V}}  \longleftrightarrow \mathbf{V}_\mathcal{A}
\end{equation*}

Here $\mathbf{V}_\mathcal{A}$ is a {\em multi-sorted structure}, each sort of which is a {\em Zariski geometry} (a notion to be defined in section \ref{Zariski}). An essential part of the multi-sorted structure $\mathbf{V}_\mathcal{A}$ are morphisms between sorts functorially agreeing with embeddings between certain sublagebras of $\mathcal{A}$, which makes the left-to-right  arrow a functor between a category of those subalgebras $A$ and sorts $\mathbf{V}_A$ of $\mathbf{V}_\mathcal{A}$.  This functor, in fact, defines a quite rich sheaf over the category of subalgebras. This makes an interesting point of contact of our approach with the topos-theoretic approach to foundations of physics suggested by C. Isham and J. Butterfield and developed by C. Isham, A. D\"oring and others, see \cite{ID}. In this approach $\mathbf{V}_\mathcal{A}$ is supposed to be a topos.\\ 

From \cite{bo4} we have the duality between rational Weyl algebras $A$ and corresponding (quantum) Zariski geometries $\mathbf{V}_A$. The structure $\mathbf{V}_A$ encapsulates the representation theory of $A$. In \cite{bo3} these ideas are used to extend the duality on algebras $A$ approximated by rational Weyl algebras. This requires to develop a notion of approximation on the side of structures $\mathbf{V}_{A}$ which is also done there. In this paper we are going to summarize the constructions in the referred paper and discuss their relevance for physics. More precisely, it will be shown that those constructions yield that quantum mechanics is represented in the limit of one particular module (called principal). This limit structure is called the space of states. \\

Some words about the organization of the paper are in order. In section \ref{Zariski} we give a brief summary on Zariski geometries for the convenience of the reader. In sections \ref{Weyl} and \ref{physics} we review some of the results in \cite{bo3} on rational Weyl algebras,algebraic Hilbert spaces and the structural approximation procedure which are relevant for the physical point of view. We will mention the sort of computations and physical problems that can be tackled by the using the theoretical tools described in these sections. Finally, in section \ref{applications} we discuss some open questions and further developments.  

\section{Zariski geometries} \label{Zariski} 

Zariski geometries were introduced in \cite{hrbo} as a generalization of Zariski topologies, a well known concept in algebraic geometry, in order to study the hierarchy of stable structures by introducing a topological ingredient in logic.\\

A structure $\mathrm{M}$ is a pair $(M,L)$ where $M$ is a set called the {\it universe} of $\mathrm{M}$ and a family $L$ of relations in $M$ (equivalently, subsets of Cartesian powers of $M$) which is called the {\it language} for $\mathrm{M}$. The structure is called {\it topological} when $L$ generates a topology on $M^n$ (a basis of closed sets of the topology), for every $n$, with the conditions that the projections $\mathrm{pr}_{i_1,...,i_k} : M^n \longrightarrow M^k$ are continuous. One of the desired things to do is to assign a dimension (with nice properties) to every closed set $C$ and its projections. In this case we say that $\mathrm{M}$ is a topological structure with good dimension notion. For those structures is possible to define a notion of {\it presmoothness} in the following terms: An open irreducible set $U$ is said to be presmooth if, for any irreducible relatively closed subsets $S_1, S_2 \subset U$, and any irreducible component $S_0$ of the intersection $S_1 \cap S_2$ we have that 

\begin{equation*}
\mathrm{dim} S_0 \geq \mathrm{dim} S_1 + \mathrm{dim} S_2 - \mathrm{dim} U
\end{equation*}

Taken together the assumptions above we arrive to the definition of {\bf Zariski geometry}. Note that in the literature, see for example \cite{bo}, Zariski geometries may appear with the adjective {\it Noetherian} or {\it Analytic}. However, in this text that distinction is not relevant, so we just avoid it. The basic examples of presmooth Zariski geometries come from algebraic geometry. For instance, let $M$ be the set of $\mathbb{F}$-points of a smooth algebraic variety over an algebraically closed field $\mathbb{F}$ and for $L$ , take the family of Zariski subsets of $M^n$ for all $n$. The dimension is the Krull dimension. \\

In \cite{hrbo} was proven an important classification theorem for a one dimensional non-linear Zariski geometry $\mathrm{M}$. It says that there exists a quasi-projective algebraic curve $C_{M}(\mathbb{F})$ over an algebraic closed field $\mathbb{F}$ and a surjective map $p: \mathrm{M} \longrightarrow C_M(\mathbb{F})$, such that for every closed $S\subseteq M^n$, the image $p(S)$ is Zariski closed in $C_{M}^n$ (in the sense of algebraic geometry) and if $\hat{S} \subseteq C_M^n$ is Zariski closed then $p^{-1}(\hat{S})$ is a closed subset of $M^n$ (in the sense of Zariski geometry). \\

In other words, the theorem referred above asserts that in the one dimensional case a non-linear Zariski geometry is \emph{almost} an algebraic curve. We want to note that the adverb \emph{almost} is really relevant, since examples of non-classical Zariski geometries (in the sense that they are not definable in algebraically closed field) are known, see \cite{bo}. The links with non-commutative geometry arise in the study of these non-classical geometries.\\

In case when the Zariski geometry $\mathrm{M}$  is associated with an algebra which is a $C^*$-algebra, one can apply the functor    $$\mathrm{M}\mapsto \mathrm{M}_\mathbb{R}$$
which ``cuts out'' the {\bf real part of}  the structure $\mathrm{M}.$  E.g. for the commutative algebra
over $\mathbb{C}$ generated by two invertible elements $X$ and $Y$ the associated geometry  $\mathrm{M}$ is just
the complex algebraic variety $\mathbb{C}^\times \times \mathbb{C}^\times$ (the complex algebraic 2-torus), while  $\mathrm{M}_\mathbb{R}$ is $S^1\times S^1,$ the square of the unit circle $S^1.$

An interesting example of  a non-commutative $\mathrm{M}$ with  a discrete  real part is discussed in \cite{SSZ}.
\\

These are the main lessons that one may learn from the study of non-classical Zariski geometries which are part of the core of this paper: 
\begin{itemize}
\item The class of (non-classical) Zariski geometries extends algebraic geometry over algebraically closed fields into the domain of non-commutative and quantum geometry. \\  
\item The non-commutative coordinate algebras for Zariski geometries emerge essentially for the same reasons as they did in quantum physics. \\
\item Zariski geometries serve as a geometric counterpart in the duality algebra-geometry (syntax-semantics) for a large class of quantum algebras, extending that duality to the non-commutative realm.

\item $C^*$-algebras correspond to  the real parts of a Zariski geometries. 
\end{itemize}

\section{Rational Weyl algebras and algebraic Hilbert spaces} \label{Weyl}

In this paper we will be interested in the Heisenberg-Weyl algebra $\mathbb{A}$ generated by $P$ and $Q$ the ``co-ordinates''  of one-dimensional quantum mechanics\footnote{Much of the discussion here also applies to the general case of the $n$-th Heisenberg-Weyl algebra $\mathbb{A}^{(n)}$ ``generated by
self-adjoint operators'' $P_1,\ldots, P_n, Q_1,\ldots, Q_n,$ but we do not need to work in such generality for the purposes of this paper.}, with  the canonical commutation relation  
\begin{equation*}\label{ccr} 
QP-PQ= i \hbar.
\end{equation*}

The meaning of ``generated'' depends on the choice of topology on the algebra, and is one of the sources of troubles with this algebra.   

The representation theory of $\mathbb{A}$ is notoriously complicated, in particular due to the fact that the algebra can not be represented as an algebra of bounded operators on a Hilbert space. \\

As it was suggested by Hermann Weyl and following the Stone--von Neumann theorem we may consider the representation theory of algebras generated by the {\em Weyl operators} which can be formally defined as 

\begin{equation*}\label{df}
U^{a}=\exp  iaQ,\ \ V^{b}=\exp ibP
\end{equation*}
for  $a ,b \in \mathbb{R}$. These  are unitary (and so bounded) operators if $P$ and $Q$ are self-adjoint, and the following commutation relation holds: 

\begin{equation*} \label{ccr1} 
U^{a}V^{b}=qV^{b}U^{a} 
\end{equation*}
where  $q=\exp {iab\hbar}$. \\

The complex $C^*$-algebra generated by Weyl operators $U^{\pm a}, V^{\pm b}$
$$A(a,b)=\mathbb{C}[U^{\pm a},V^{\pm b}]$$ is called (in this paper) a Weyl algebra. Now, one can say that $\mathbb{A}$ can be fully replaced by the entirety of its Weyl subalgebras  $A(a,b)$ which have a good Hilbert space representation theory. \\

By rescaling we may assume that $\frac{\hbar}{2\pi}$ is a rational number.
 Now 
when $a,b$ are rational numbers, the corresponding Weyl algebras have very nice finite-dimensional representations since the multipliers $q=e^{iab\hbar}$ are roots of unity (of order
$N=N(a,b),$  the denominator of the rational number $\frac{ab\hbar}{2\pi}.$)
 These quantum algebras at roots of unity and their representation theory are very well understood (see e.g \cite{Brown&Goodearl}), we call them {\bf rational Weyl algebras}. A crucial idea of \cite{bo3} is that in the same way that rational points approximate points of the real line we should have that rational Weyl algebras approximate the full Heisenberg-Weyl algebra. We will make this idea more precise below.  \\

Let us consider the category $\mathcal{A}_{fin}$ of all rational Weyl algebras with canonical embeddings as morphisms. We have that an algebra $A= A(a,b) \in \mathcal{A}_{fin}$ is an affine prime algebra over $\mathbb{C}$ generated over its centre $Z(A)$ as a module. In fact, we have that both $A$ and $Z(A)$ are prime Noetherian principal ideals ring. Every maximal ideal of $A$ is of the form $\tilde{\alpha} = A\alpha$ for $\alpha$ in $\mathrm{Spec}(Z(A))$ and an isomorphism type of an irreducible module $\mathrm{V}$ is fully determined by $\alpha$ such that $\mathrm{Ann}(\mathrm{V})= \tilde{\alpha}.$ We call this $A$-module $\mathrm{V}_A(\alpha).$ Note that the dimension of $\mathrm{V}_A(\alpha)$ over $\mathbb{C}$ is $N$ for every $\alpha\in \mathrm{Spec}(Z(A)).$\\

The next standard step in the theory of quantum algebras at root of unity allows one to choose a {\em canonical $U^a$-basis $\mathbf{u}(\alpha),$} which is of size $N.$ However
given $\alpha$ a canonical $U^a$-basis $\mathbf{u}(\alpha)$ is determined only up to the finite group $\Gamma_A$ of symmetries corresponding to the Galois symmetries of roots of unity of order $N,$ so we associate to $\alpha$ a finite family of bases $\mathbf{u}(\alpha)$. The important point is that a choice of a basis  $\mathbf{u}(\alpha)$ allows us to introduce the unique inner product in $\mathrm{V}_A(\alpha)$ with respect to which $\mathbf{u}(\alpha)$ is an orthonormal basis. A change to another canonical basis does not effect inner product since the transformations of $\Gamma_A$ are given by unitary matrices. \\

Analogously we may choose  canonical $V^b$-bases and more general canonical $U^{an}V^{bm}$-bases, for $m,n\in \mathbb{Z}.$ However, any such base can be obtained from  $\mathbf{u}(\alpha)$ by applying a unitary transition matrix with coefficients in the cyclotomic field $\mathbb{Q}(q).$ The structure on $\mathrm{V}_A(\alpha)$ endowed with the canonical bases will be called an {\bf algebraic-Hilbert space} and denoted by $\mathbf{V}_A(\alpha)$. \\

The described picture is illustrated in figure 1 as a fibre-bundle $\mathbf{V}_A$ over a torus. The points of the torus are elements $\alpha$ of the spectrum $\mathrm{Spec}(Z(A)).$ Over each such $\alpha$ we set a finite structure which consists of all the canonical bases  $\mathbf{u}(\alpha).$  Knowing a canonical base is the same as knowing the algebraic-Hilbert space $\mathbf{V}_A(\alpha),$ so just as well we may think of the fibre over $\alpha$ as the algebraic-Hilbert space $\mathbf{V}_A(\alpha).$ The fibres are acted upon by elements of $A.$ \\

The structure $\mathbf{V}_A$ over the torus resembles a vector bundle as defined and studied in algebraic geometry. However, one of the main results of \cite{bo4} is that this structure in full {\bf can not be defined in terms of commutative algebraic geometry}. Yet the more general context of a Zariski geometry is applicable, as explained in the same paper, which  allows to treat $\mathbf{V}_A$ as a structure with a (generalised) Zariski topology on it. In particular, we can distinguish between maps which are {\em Zariski-regular} and those which are not. 
 
\begin{center}
\begin{tikzpicture} [scale=0.8]
\draw[ultra thick] (0,4.5) arc (270:90:0.2);
\draw[ultra thick, <-] (0.3,4.5)--(0,4.5);
\draw[ultra thick] (3,4.5) arc (270:90:0.2);
\draw[ultra thick, <-] (3.3,4.5)--(3,4.5);
\draw[ultra thick] (-3,4.5) arc (270:90:0.2);
\draw[ultra thick, <-] (-2.7,4.5)--(-3,4.5);

\node at (0,6.5) {$A$};
\node at (0,3.5) {$\mathbf{V}_A(\alpha_j)$};
\node at (3,3.5) {$\mathbf{V}_A(\alpha_k)$};
\node at (-3,3.5) {$\mathbf{V}_A(\alpha_i)$};
\node at (0,1) {$\alpha_j$};
\node at (3,1) {$\alpha_k$};
\node at (-3,1) {$\alpha_i$};
\node at (0,-3.5) {Figure 1};
\node at (6,0) {$\mathrm{Spec}(Z(A))$};

\draw[<-] (0,5) -- (0,6);
\draw[<-] (3,5) -- (0,6);
\draw[<-] (-3,5) -- (0,6);

\draw[gray, thick] (0,1.5) -- (1.2,4);
\draw[gray, thick] (0,1.5) -- (-1.2,4);
\draw[gray, thick] (3,1.5) -- (4.2,4);
\draw[gray, thick] (3,1.5) -- (1.8,4);

\draw[gray, thick] (-3,1.5) -- (-1.8,4);
\draw[gray, thick] (-3,1.5) -- (-4.2,4);

\filldraw [black] (0,1.5) circle (2pt);
\filldraw [black] (3,1.5) circle (2pt);
\filldraw [black] (-3,1.5) circle (2pt);
\filldraw [black] (0,6) circle (2pt);

\draw (-4,0) .. controls (-4,2) and (-1.5,2.5) .. (0,2.5);
\draw[xscale=-1] (-4,0) .. controls (-4,2) and (-1.5,2.5) .. (0,2.5);
\draw[rotate=180] (-4,0) .. controls (-4,2) and (-1.5,2.5) .. (0,2.5);
\draw[yscale=-1] (-4,0) .. controls (-4,2) and (-1.5,2.5) .. (0,2.5);
\draw (-2,0.2) .. controls (-1.5,-0.3) and (-1,-0.5) .. (0,-.5) .. controls (1,-0.5) and (1.5,-0.3) .. (2,0.2);
\draw (-1.75,0) .. controls (-1.5,0.3) and (-1,0.5) .. (0,.5) .. controls (1,0.5) and (1.5,0.3) .. (1.75,0);

\end{tikzpicture}
\end{center}

\section{From algebraic-Hilbert spaces to quantum mechanics via sheaf of Zariski geometries} \label{physics}

In the many earlier treatments of quantum physics over roots of unity (see e.g. \cite{SN}) the research is concentrated on one $A$ (with large enough $N$) and one particular $A$-module $ \mathbf{V}_A(\alpha)$ in which the $U^a$-  and
$V^b$-eigenvalues are roots of unity of order $N.$  This module, which we call the {\bf principal module}
and denote $\mathbf{V}_A(\bf{1}),$ plays an important role in our study as well (see below).
 Nevertheless, studying this module alone one loses the information encoded in the full algebra $A$ and only ``sees'' the quotient of $A$ modulo the annihilator of  $\mathrm{V}_A(\bf{1}).$ For a similar reason, in order to understand the semantics of  the whole algebra $\mathbb{A},$  we need to take into account all the $A  \in \mathcal{A}_{fin},$ so the whole category $\mathcal{A}_{fin}.$
 \\

Following the section above we can construct a category $\mathcal{V}_{fin}$ of Zariski geometries which consists of objects $\mathbf{V}_{A}$, for $A \in \mathcal{A}_{fin}$ and morphisms $p_{AB}:$ {\bf V}$_{B} \longrightarrow$ $\mathbf{V}_{A}$ , $B \subseteq A$. 
This category is a sheaf on the category $\mathcal{A}.$ 
In model-theoretic setting, $\mathcal{V}_{fin}$ can be seen as a multisorted structure.\\

As it was mentioned in the introduction in \cite{bo4} it was  established the duality $$A \longleftrightarrow {V}_A$$
between quantum algebras $A$ at roots of unity (rational Weyl algebras in our case) and corresponding Zariski geometries $\mathbf{V}_A$, which extends the classical duality between commutative affine algebras and affine algebraic varieties.  The new step implemented in \cite{bo3} is to extend the duality on algebras approximated by rational Weyl algebras. In order to do that we need to extend the categories $\mathcal{A}_{fin}$ and $\mathcal{V}_{fin}$ by adding some limit objects. \\

The {\bf universal limit object} for $\mathcal{A}_{fin}$ should be defined as the full Heisenberg-Weyl algebra
$\mathbb{A}\supseteq \bigcup _{A \in \mathcal{A}_{fin}} A.$
 The important thing to note is that the center $Z(\mathbb{A}) = \bigcap_{A \in \mathcal{A}} \subseteq \mathbb{C}$ of this limit object is trivial, so $\mathrm{Spec}(Z(\mathbb{A})) = \lbrace {\bf 1} \rbrace$. Therefore, in $\mathcal{V}_{fin}$  in the limit we should have a fibre-bundle $\mathbf{V}_{\mathbb{A}}({\bf 1})$ over a point as it is illustrated in figure 2.

\begin{center}
\begin{tikzpicture} [scale=0.5]

\draw[ultra thick] (0,4.5) arc (270:90:0.4);
\draw[ultra thick, <-] (0.5,4.5)--(0,4.5);

\node at (0,6) {$\mathbb{A}$};

\filldraw [black] (0,0) circle (5pt);
\draw[gray, thick] (0,0) -- (-6,5);
\draw[gray, thick] (0,0) -- (6,5);

\node at (0,3){$\mathbf{V}_{\mathbb{A}}({\bf 1})$};
\node at (0,-1){$\mathrm{Spec}(Z(\mathbb{A})) = \lbrace {\bf 1} \rbrace$};
\node at (0,-2.5){Figure 2};

\end{tikzpicture}
\end{center}

Comparing with figure 1 we can say that  $\mathrm{Spec}(Z(\mathbb{A}))$ goes down to  $\lbrace {\bf 1} \rbrace$ as $A$ goes up to the universal object $\mathbb{A}.$ The collapse of  $\mathrm{Spec}(Z(\mathbb{A}))$ is compensated by the blow-up of $\mathbf{V}_{\mathbb{A}}({\bf 1})$ to an infinite dimensional vector space. However, the full structure on  $\mathbf{V}_{\mathbb{A}}({\bf 1})$ is more involved and its construction via a limit procedure called the structural approximation (see \cite{bo2}) is a major result of the work. \\

Note that the object $\mathbb{A}$ does not belong to $\mathcal{A}_{fin}$ and our first aim is to approximate it by an object  $\tilde{\mathbb{A}}$ which in a naive sense is a limit of objects of $\mathcal{A}_{fin}.$

We will start with the family   $\mathcal{A}_{fin}$ of all rational Weyl algebras and the Fr\'{e}chet filter $\mathcal{D}$  with respect to the partial ordering, that is for each $A$ the set $\lbrace B \in \mathcal{A}_{fin}   : B \subset A \rbrace$ is in  $ \mathcal{D}$.  Define $\tilde{\mathbb{A}}$ to be 
the ultraproduct of the algebras $A$ in  $\mathcal{A}_{fin}$ modulo the ultrafilter  $\mathcal{D}.$
One can identify this $\tilde{\mathbb{A}}$
 as being ``pseudo-finitely generated'' by a pair of operators $U^{\frac{1}{\mu}}$ and $V^{\frac{1}{\nu}}$, for non-standard integers $\mu$ and $\nu.$
 By construction the integers  $\mu$ and $\nu$ have a property of
   being divisible by all standard integers.  This property is equivalent to the fact that  $\tilde{\mathbb{A}}$ {\em is an upper bound for  $\mathcal{A}_{fin}$}, i.e.
   $A \subset \tilde{\mathbb{A}}$ for any rational $A$. 
   Now, we take the corresponding ultraproduct modulo $\mathcal{D}$ of Zariski geometries $\mathbf{V}_{A}(\mathbf{1})$ and denote it $\mathbf{V}_{\tilde{\mathbb{A}}}(\mathbf{1})$. \\

The full limit structure $\mathbf{V}_{\mathbb{A}}(\mathbf{1})$ is the  image of $\mathbf{V}_{\tilde{\mathbb{A}}}(\mathbf{1})$ by a {\it homomorphism} named $\mathbf{lim},$

\begin{equation*}
\mathbf{lim} : \mathbf{V}_{\tilde{\mathbb{A}}}(\mathbf{1}) \longrightarrow \mathbf{V}_{\mathbb{A}}(\mathbf{1}).
\end{equation*}

Here homomorphism means that the map preserves basic relations and operations of the language. 
In particular, the images of Zariski closed subsets of (the cartesian powers of)  $\mathbf{V}_{\tilde{\mathbb{A}}}(\mathbf{1})$  give rise to a topology on $\mathbf{V}_{\mathbb{A}}(\mathbf{1})$ and Zariski regular maps become well-behaved continuous maps on the image. Pseudo-finite summation formulas over Zariski regular maps in $\mathbf{V}_{\tilde{\mathbb{A}}}(\mathbf{1})$ translate into integrals over the corresponding maps
in $\mathbf{V}_{\mathbb{A}}(\mathbf{1}).$

This being said we must add that there is no claim that the  image of every Zariski regular map 
is well-behaved in $\mathbf{V}_{\mathbb{A}}(\mathbf{1}).$ Finding out which ones are is a major problem for the future work. \\

At this point it is important to explain the choice of basic relations and operations of the language. A seemingly obvious choice of basic operations given by $U^a$ and $V^b$ is problematic because these are not ``globally'' defined: $U^a$ and $V^b$ make only sense for $\mathbf{V}_A,$ such that $A(a,b)\subseteq A,$ that is for $A$ at a high enough level of 
  $\mathcal{A}_{fin}.$ Instead we define in each $A(a,b)$
  $$Q:= \frac{U^a-U^{-a}}{2ia} \mbox{ and }P:= \frac{V^b-V^{-b}}{2ib} .$$  
These are inter-definable  with $U^a$ and $V^b$ in each $\mathbf{V}_{A(a,b)}$ (by formulae depending on the rational parameters $a$ and $b$), but the advantage of using $Q$ and $P$ for basic symbols is that these make sense everywhere in $\mathcal{A}_{fin}$ as well as in $\mathbf{V}_{\tilde{\mathbb{A}}}(\mathbf{1})$ and $\mathbf{V}_{\mathbb{A}}(\mathbf{1}).$ This, we believe, what makes the true difference between {\bf observables} and the other operators and 
relations in the suggested interpretation of quantum mechanics.

It is important to note that the $a$ and $b$ in our definition
of $P$ and $Q$ become infinitesimals in $\tilde{\mathbb{A}}$ and
thus the limit values of the expressions give us limit elements of $\mathbb{A}$ satisfying the canonical commutation relation.
\\
 
It turns out that the properties and the invariants of $\mathbf{V}_{\tilde{\mathbb{A}}}(\mathbf{1})$ and the limit object $\mathbf{V}_{\mathbb{A}}(\mathbf{1})$ depend essentially on the ratio $\mathrm{h} = \frac{\mu}{\nu}$. It is convenient to assume that  $\mathrm{h}$ is a rational number or, more generally, that $\mathrm{h}$ is a non-standard finite rational number (so then
$\frac{\mathrm{h}}{\nu}$ is infinitesimal).
 \\

In terms of physics the  main object 
$\mathbf{V}_{\mathbb{A}}(\bf{1}_{\mathbb{R}})$ (or rather a certain collection of norm 1 elements of it) can be thought of as  {\bf the space of  states}. This space is a substitute for the Hilbert space of states of quantum mechanics (corresponding to the {\it continuous limit} version of universe). On the other hand, the corresponding set of states of $\mathbf{V}_{\tilde{\mathbb{A}}}(\bf{1})$ is  pseudo-finite (i.e. its size is a non-standard integer) and can be thought of  as the 
actual ``huge finite universe'' while the space of  states of $\mathbf{V}_{\mathbb{A}}(\bf{1}_{\mathbb{R}})$ is only its ``observable image''. 
\\

One of the main important points of the construction described above is that the approximation procedure $\mathbf{lim}$ suggests a rigorous interpretation for passing from discrete models, i.e. the finite dimensional spaces $\mathbf{V}_A({\bf 1})$ representing the discrete Dirac calculus (as in \cite{fis} or \cite{SN}), to the continuous limit. \\

The scheme for practical calculations (rigorous Dirac calculus) can be described as follows:\\

1. Rewrite a formula over $\mathbf{V}_{\mathbb{A}}(\bf{1}_{\mathbb{R}})$ in terms of Zariski-regular pseudo-finite sums and products over $\mathbf{V}_{\tilde{\mathbb{A}}}(\bf{1})$. \\

2. Calculate in $\mathbf{V}_{\tilde{\mathbb{A}}}(\bf{1})$ using, for example, the Gauss quadratic sums formula.\\

3. Apply $\mathbf{lim}$ and check that  the result in terms of the standard real numbers  is well-defined. \\

A straightforward application of this scheme prove

\begin{itemize}
\item the canonical commutation relation holds in the space of states: $$\mathrm{QP}- \mathrm{PQ}=i\hbar;$$
\item  the Feynman propagator $\langle x|e^{-i\frac{H}{\hbar}t}y\rangle$ is well-defined 
and coincides with the well-known formulae for $H=\frac{\mathrm{P}^2}{2}$ (free particle Hamiltonian) and $H=\frac{\mathrm{P}^2+\mathrm{Q}^2}{2}$ (harmonic oscillator);
\item for  $H=\frac{\mathrm{P}^2+\mathrm{Q}^2}{2},$
$$\int_\mathbb{R} \langle x|e^{-i\frac{H}{\hbar}t}x\rangle dx=\frac{1}{i\sin \frac{t}{2}},$$
the trace of the time evolution operator for the Harmonic oscillator formula.
\end{itemize}

For details we refer the reader to \cite{bo3}.

\section{Questions and further developments}  \label{applications}

In the construction described in the previous section there are two important questions which haven been left open and deserves further study. These are: 

\begin{itemize}

\item How does the structure $\mathbf{V}_{\tilde{\mathbb{A}}}$ depend on the choice of the Fr\'{e}chet ultrafilter $\mathcal{D}$?  
\item Does the construction of the space of  states result in a unique object? 

\end{itemize}

We hope that the {\em elementary theory} (a model-theoretic notion) of $\mathbf{V}_{\tilde{\mathbb{A}}}$ does not depend on  $\mathcal{D}$ and
that  the state of space is determined uniquely. 
However, even in the case of not getting the ``desired'' answers it would be interesting to see then how the choice of an ultrafilter ``affects'' the structure $\mathbf{V}_{\tilde{\mathbb{A}}}$ (and how it varies when the ultrafilter changes) and what are the relations between the different spaces of states, if any. It would be also interesting to see what is obtained in these situations in terms of concrete computations. These questions are important for both the physics and model-theoretic point of view.\\


An important task to be undertaken consists in extending the approach we have described in order to make more computations and cover a wider context. In this direction, we hope that our approach can be used to formalise path integrals in the quantum mechanics context and it could be extended to cover free quantum fields and related stuff. We plan to do this somewhere else. \\

Finally, we would like to mention a subtle difference between the numerical systems naturally related to the continuous limit version of the space, $\mathbf{V}_{{\mathbb{A}}}(\mathbf{1}_\mathbb{R}),$
and the ``huge finite universe'' one, $\mathbf{V}_{\tilde{\mathbb{A}}}(\mathbf{1}).$  While in the first measurements are in terms of the field of  real numbers $\mathbb{R},$   the second is based on $\tilde{\mathbb{Z}}/(\mu),$
the ring of non-standard integers modulo an infinite non-standard number $\mu$ described above. The latter object familiar to model-theorists exhibits many number-theoretic properties akin to the properties of standard integers, but is in many ways different.  

The study of $\tilde{\mathbb{Z}}/(\mu)$ in the context of the suggested semantics might shed a light on  a  series of intriguing connections between number theoretic  and physics phenomena discussed e.g. in \cite{mar}.  




\end{document}